\documentclass[epsfig]{aa}
\usepackage[dvips]{graphics}

\begin{document}

   \thesaurus{03     
              (11.01.2;
               13.20.1;
               13.09.1;
               13.25.2;
               12.04.2)}
   \title{Sub--mm and X--ray background: two unrelated phenomena?}


   \author{P. Severgnini
          \inst{1}
	\and
	  R. Maiolino, \inst{2}
	  M. Salvati,   \inst{2}
	  D. Axon, \inst{3}
	  A. Cimatti, \inst{2} 
	  F. Fiore,  \inst{4}
	  R. Gilli, \inst{1,5}
	  F. La Franca,  \inst{6}
	  A. Marconi,  \inst{2}
	  G. Matt,  \inst{6}
	  G. Risaliti  \inst{1}
	    \and
	  C. Vignali  \inst{7,8}    
          }

   \offprints{P. Severgnini, paolas@arcetri.astro.it}

   \institute{$^1$ Dipartimento di Astronomia, Universit\`a di Firenze,
    L.go E. Fermi 5, I--50125, Firenze, Italy\\
              $^2$ Osservatorio Astrofisico di Arcetri,
    L.go E. Fermi 5, I--50125, Firenze, Italy \\
	      $^3$ Department of Physical Sciences, University of Hertfordshire
    College Lane, Hatfield, Herts AL10 8 AB, UK\\
	      $^4$ Osservatorio Astronomico di Roma, Via Frascati 33, I--00044 Monteporzio
   Catone, Italy \\
	      $^5$ Astrophysikalisches Institut Potsdam,
   an der Sternwarte 16, 14482--Postsdam, Germany \\
	      $^6$ Dipartimento di Fisica, Universit\`a degli Studi ``Roma Tre''
   Via della Vasca Navale 84, I--00146, Roma, Italy \\
	      $^7$ Dipartimento di Astronomia, Universit\`a di Bologna,
   Via Ranzani 1, I--40127 Bologna, Italy \\ 	
	      $^8$ Osservatorio Astronomico di Bologna,
   Via Ranzani 1, I--40127 Bologna, Italy 
            } 

   \date{Received ......; accepted ..........}

 \titlerunning{Sub--mm  and X--ray background}
 \authorrunning{P. Severgnini et al.}
 \maketitle

\maketitle

\begin{abstract}
Obscured AGNs are thought to contribute a large fraction
of the hard X-ray background (2--10 keV), and have also been 
proposed as the powerhouse of a fraction of 
the SCUBA sources which make most of 
the background at 850$\mu$m, thus providing a link between the 
two spectral windows.
We have tackled this issue by comparing data at 2--10 keV and at 850$\mu$m
for a sample of 34 
sources at fluxes (or limiting fluxes) which resolve
most of the background in the two bands. We present here
new SCUBA observations, and new correlations between separate data
sets retrieved from the literature. Similar correlations presented
by others are added for completeness.
None of the 11 hard X-ray (2--10 keV) sources has a
counterpart at 850 $\mu$m, with the exception of a Chandra source
in the SSA13 field,
which is a candidate type 2, heavily absorbed QSO at high redshift.
The ratios $\rm F_{850\mu m}/F_{5keV}$ (mostly upper limits) of
the X-ray sources are significantly lower
than the value observed for the cosmic background. In particular, we obtain
that 2--10 keV sources brighter than $\rm 10^{-15}~erg~s^{-1}cm^{-2}$, which
make at least 75\% of the background in this band, contribute for less
than 7\% to the submillimeter background.
Out of the 24 SCUBA sources, 23 are undetected by
Chandra. The ratios $\rm F_{850\mu m}/F_{5keV}$ (mostly lower limits)
of these SCUBA sources indicate that most of them must be powered either
by starburst activity, or by an AGN which is obscured by a column 
$\rm N_H > 10^{25}~cm^{-2}$, with a reflection efficiency in the
hard X~rays significantly lower than 1\% in most cases.
However, AGNs of this type could not contribute significantly to the
2--10 keV background.

   \end{abstract}

%
\section{Introduction}
Active Galactic Nuclei (AGNs) are thought to produce most of the X-ray 
background (XRB) in the 1--100 keV energy range. 
Deep ROSAT observations resolved 80$\%$ of the XRB 
around 1 keV into discrete sources (Hasinger et al. \cite{Has}), most of which were
identified with low absorption type~1 AGNs
(Schmidt et al. \cite{Sch}). 
However, the energy density of the XRB has a 
maximum around 30 keV, where type 1 AGNs cannot give a dominant 
contribution due to their relatively soft spectra.  
Following the original suggestion of Setti \& Woltjer (\cite{Set:Wol}),
the hard XRB is commonly explained with the superposed emission of a large 
population of highly obscured, type 2 AGNs (eg.
Comastri et al.  \cite{Com}). Recently, deep Chandra observations
in the 2--10 keV range, down to a limiting flux of
$\rm 2.5\times 10^{-15}~erg~s^{-1}~cm^{-2}$ (Mushotzky et al. 2000, hereafter
M00), resolved 75\% of the XRB in this spectral range if the background
measured by BeppoSAX is adopted, or an even
larger fraction if previous background
estimates are used (Vecchi et al. 1999 and references therein).
So far, optical identifications have been biased in favour of
optically bright sources, i.e. broad line
unobscured QSOs and Seyfert 1 galaxies, however some of the counterparts
do show indications of obscured AGNs
[M00, Fiore et al. 2000a, Brandt et al. 2000,
Fabian et al. 2000 (hereafter F00), Hornschemeier et al. 2000
(hereafter H00)].

If obscured AGNs are actually the major contributors to the hard X-ray
background, a
large fraction of their optical-UV energy must be re-radiated at longer
wavelengths, from the near to the far
infrared and submillimeter (sub-mm) regions,
as confirmed by the observations of nearby heavily absorbed 
AGNs (eg. Vignati et al. \cite{Vig}). 
The shape of the infrared spectrum of such sources is uncertain, but the 
integrated luminosity can be estimated from the X-ray luminosity. One
finds that,
if the reprocessing material were sufficiently cold, the absorbed AGNs
could contribute a substantial fraction (20--50\%) of the submillimeter
background (Almaini et al. 1999).

However, the nature of the energy
source in powerful IR and sub-mm galaxies is still matter of debate. Recent ISO
results suggest that most of them are powered by vigorous star
formation (eg. Genzel et al. 1998, Lutz et al. 1998).
If starbursts were the dominant contributors to the high-z SCUBA sources,
which make most of the sub-mm background
(Blain et al. 2000, Barger et al. 1999a, hereafter B99),
this would have important implications on the star formation history of the
Universe.
On the other hand,
several of the high-z SCUBA sources appear to host
an AGN, but it is not clear to what extent the latter contributes to their
sub-mm flux (Ivison et al. 1999). Constraints on the
hard X-ray emission of
the SCUBA sources should help to tackle this issue. 

With the aim of investigating the relation between the sub-mm and
the 2--10 keV hard X-ray backgrounds
we have started an observing program
with SCUBA, the Sub-mm Common 
User Bolometer Array (Holland et al. \cite{Hol})
at the James Clark Maxwell Telescope (JCMT), of a subsample
of the sources detected by BeppoSAX in the HELLAS survey. This
survey has resolved about $30\%$ of the 5--10 keV background into
discrete sources at a flux limit of $\sim$ 5$\times$10$^{-14}$
erg cm$^{-2}$ s$^{-1}$ (Fiore et al. 1999, 2000b). Here we present
preliminary results coming from these observations,  and combine these
with 2--10 keV and
sub-mm data of other fields.
In particular, 
we also include in this study the results from F00 on the
cross correlation of Chandra and SCUBA data of two lensing clusters,
the cross correlation performed by us
of the deeper Chandra observation
of SSA13 (M00) with SCUBA data 
of a sub-area (B99), and the results of the Chandra and SCUBA observations
on the HDF North reported by H00, with some additional correlations in the
same field performed by us.


\section{Observations and  Data Reduction}

The objects included in our program are a subsample of the optically
identified HELLAS sources. Higher priority was given to sources showing
evidence for absorption either in the X~rays or in the optical.
Relevant information
on their X-ray and optical properties are listed in Table~1.
In the case of SAXJ2302+0856 there is some ambiguity in the identification of
the counterpart, since another Emission Line galaxy pair
 is present in the BeppoSAX
errorbox. However, a ROSAT PSPC counterpart (which is most likely
 associated with
the hard X-ray source) is closer to the source observed with SCUBA, suggesting that
more than half of the X-ray emission comes from the latter.

\begin{table}
\label{pos}
\caption{HELLAS sources observed with SCUBA}
\begin{tabular}{lcccccc}
\hline
SAXJ       & $F_X^a$        & log$N_H^b$      &  R     & z
     & Class$^c$\\
\hline
\hline
0045--2515 & 3.5  & $<$23  &   17.4 &   0.111 & 1.9  \\

1054+5725   & 2.7         & $<$22         &    18.4 &  0.205 & 1.9 \\

1117+4018 & 1.3  & 22.7$\pm$0.5  &    19.9 &  1.274  & B \\

2302+0856   & 3.3  &  23.3$\pm$0.3  &    18.3 &  0.135 & ELG \\
\hline

\end{tabular}

\smallskip
$^a$ 5--10 keV flux in units of 10$^{-13}$ erg s$^{-1}$ cm$^{-2}$.
$^b$ Absorbing column density derived from the X-ray hardness ratio
by assuming an intrinsic photon index $\Gamma =1.8$ and the absorber at the
same redshift of the AGN.
$^c$ B=Blue broad line QSO, 1.9= intermediate type 1.9 AGN, ELG=Emission Line 
Galaxy.
\end{table}

The observations were made in standard point-source photometry mode 
at 450 and 850 $\mu$m, and the
data were reduced with the Starlink SURF software
(Jenness \& Lightfoot \cite{Jen:Lig}).
For each object more than one photometric observation was carried out.
Each observation was first reduced by subtracting the measurements in 
the reference beam from those in the signal beam, rejecting obvious spikes.
It was then flatfielded and corrected for atmospheric opacity.
For this purpose {\it skydips} were taken regularly to determine the sky 
opacity before and after the target observation.
Residual sky background emission was removed using the 
median of the different rings of bolometers as a background estimate.
With the extinction corrected and sky subtracted data we produced a 
final signal for each observation and then for each source we concatenated
together the individual observations producing a final coadded data set.
Two sources (SAXJ0045.7-25 and SAXJ2302+08) were observed on two different 
nights, with a very low and stable opacity. The remaining two were observed
on one night only, and one of them (SAXJ1117+40) suffered from bad opacity 
conditions. A primary calibrator (Uranus) was used in the August run, yielding
a 10\% accuracy, while a secondary calibrator (OH231.8) 
was used in January, yielding a calibration uncertainty of 20\%.


\begin{table}
\caption{Observing log and results}
\begin{tabular}{lcccccc}
\hline
Name & Date & $t_{\rm int}^a$ & $\tau_{850}^b$ &$\tau_{450}^b$ &$F_{850}^c$&$F_{450}^c$ \\
 SAXJ       & & (ks)        &                        &     & (mJy) & (mJy) \\
\hline
\hline
0045--2515 & 25-26/08/99 & 2.8 & 0.17 & 0.91 & $<$2.6 &$<$42\\
1054+5725 & 09/01/00 & 1.8 & 0.19 & 1.05 & $<$3.4 & $<$37 \\
1117+4018 & 10/01/00 & 1.4 & 0.33 & 2.7 & $<$7 &   - \\
2302+0856 & 20-24/08/99 & 3.0 & 0.24 & 1.39 &$<$2.4 & $<$31\\
\hline
\end{tabular}
\smallskip
$^a$ Integ. time. $^b$ Average opacity. $^c$
Upper limits are at 2$\sigma$.
\end{table}

\begin{figure*}[!h]
\resizebox{14truecm}{!}{\includegraphics{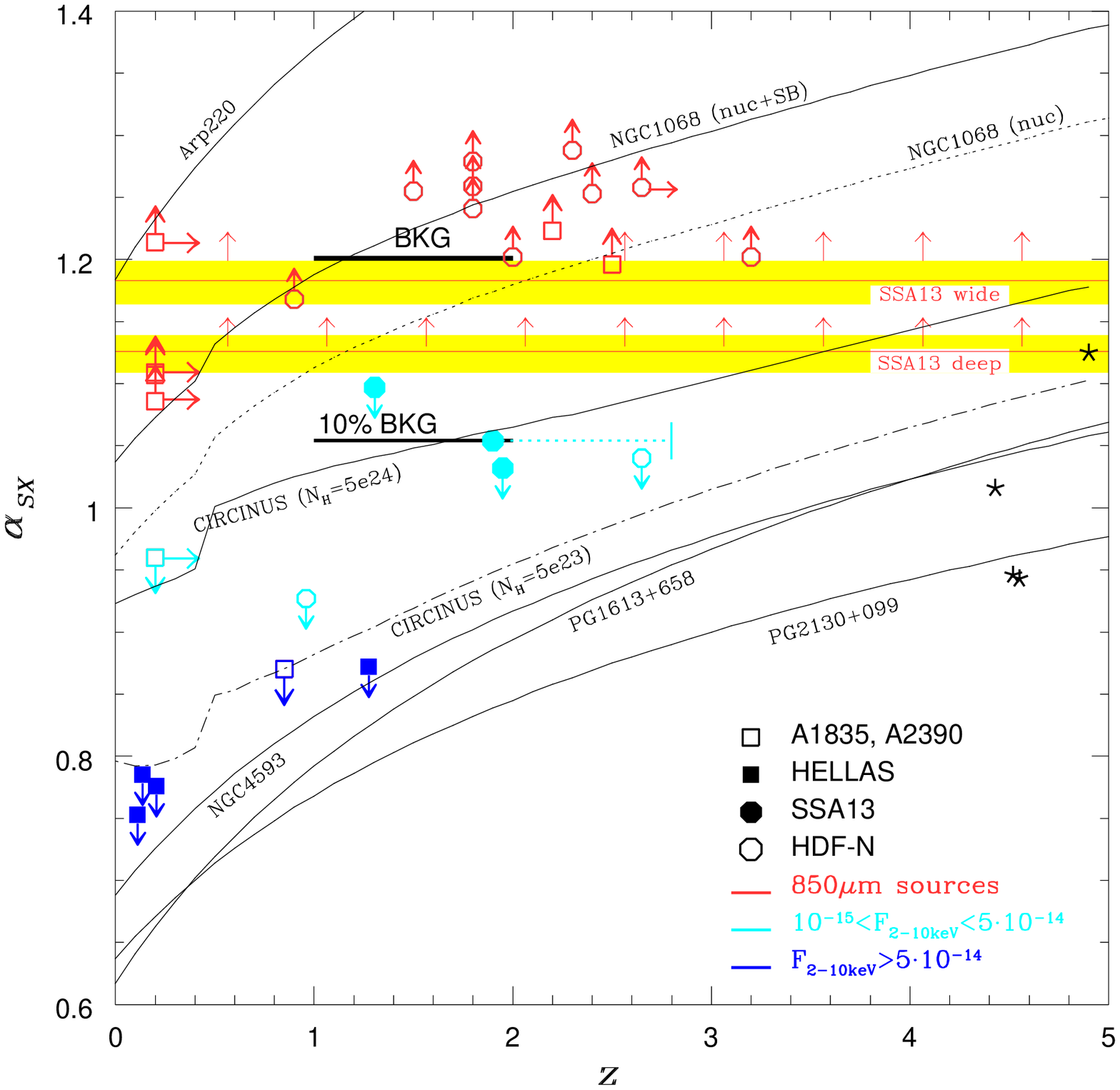}}
\caption{Distribution of the observed sub-mm to hard X-ray spectral indices
$\alpha _{SX}$ as a function of redshift.
Filled squares are the HELLAS sources observed by us.
Empty squares are the sources reported by F00.
The filled circles are Chandra hard-X sources detected by M00
 in the SSA13 field. The shaded areas give the
distribution of lower limits for the SCUBA sources detected by B99
 in the SSA13 field but undetected by Chandra. The empty circles are sources
in the HDFN reported by H00.
Dark and light blue symbols indicate 2--10 keV sources detected in different
ranges of flux (see legend), while red symbols indicate SCUBA sources
which were not detected by Chandra in the 2--10 keV band.
Templates derived from various types of known active
galaxies are shown (see text). Stars are high-z QSOs detected by SCUBA.
The thick horizontal segments give the $\alpha _{SX}$ of the cosmic
background and of sources having a factor of 10 lower sub-mm flux relative
to the 5 keV flux.}
\end{figure*}

\section {Results and cross-correlation of other surveys}

None of the four HELLAS sources in our subsample was detected. Table 2
gives the upper limits at 2$\sigma$ along with other relevant information
on the observations.
In order to put these results into perspective, we follow F00
and compute a submillimeter--to--X-ray index $\alpha_{SX}$
($\rm F_{\nu}\propto \nu^{-\alpha}$), using the 
observer-frame flux densities at 850 $\mu$m and 5 keV.
The flux densities at 5 keV have been derived from the observed 5--10 keV 
fluxes and spectral indices. In Fig.~1, which gives $\alpha_{SX}$
as a function of redshift, our results are represented by filled squares
(note that in Fig.~1 the Y-axis is inverted with respect to F00, this
is to avoid misunderstanding when discussing upper and lower limits).

The open squares are the sources of F00, modified
in accordance with the energy band used here, which were detected either at
850$\mu$m by SCUBA or in the 2--10 keV band by Chandra.

We have also cross-correlated the
deep hard X-ray survey of the Hawaii Field SSA13, carried 
out by M00
with Chandra, with the
deep ($\sigma =0.6-1.0$ mJy) small-area and, respectively, shallow 
($\sigma = 1.5-2.5$ mJy) wide-area SCUBA maps centered 
on the same field obtained by B99
at 850 $\mu$m. Three of the hard X-ray sources detected by M00
lie in the deep SCUBA area (\# 15, 18 and 21 in M00)\footnote{The
upper limits on
$\alpha _{SX}$ from the M00 hard-X sources in the {\it wide}
 SCUBA field could not be derived
because of insufficient information on the area covered by the latter.}.
Only one (\#15) out of these three sources has a
counterpart at 850 $\mu$m: it has not been spectroscopically identified, but
its colours match those of a young galaxy at z$\sim$1.9, with significant
probability at z$\sim$2.6--2.8 as well. This is also
consistent with the HST optical image (fuzzy and possibly interacting).
Object \#15 is
indicated with a filled circle in Fig.~1; a horizontal dotted bar
shows its redshift uncertainty. At this redshift, the observed flux
implies a luminosity of $L_{2-10keV} \sim 1-2\times 10^{44}$ erg~s$^{-1}$ (or
intrinsically higher if absorbed)\footnote{We assume
$\rm H_0=50$ and $\rm q_0=0.5$.}, which is not extreme, but
 in the QSO range.
Although no information is available on its X-ray spectral slope, the optical
to near-IR properties (see above) and, as we will see, the $\alpha _{SX}$
suggest that this is a type 2 QSO. Out of the other two hard X-ray sources
in the deep SCUBA area, one (\# 21)  is spectroscopically
identified with a QSO at z$=$1.3 and for the other (\# 18) we derive a
photometric redshift z$=$1.9-2.0
(upper limits with filled circle in Fig.~1).
On the other hand, none of the 9 SCUBA
sources (but the one discussed above) detected by B99, both in the
deep and shallow survey areas, has a Chandra counterpart, giving lower limits
on their $\alpha _{SX}$. Since for none of these sources the redshift is known,
the spread of the
lower limits on $\alpha _{SX}$ is shown with two yellow
shaded areas on Fig. 1
(the thin red solid lines give the mean of the lower limits).

Finally, we also report in Fig.~1 the lower limits on $\alpha _{SX}$
for the 10 SCUBA 850$\mu$m
sources which were undetected by Chandra in the HDFN by H00
(open circles).
We also derived the upper limits on $\alpha _{SX}$ for the two sources
detected by Chandra in the 2--10 keV band in the HDFN (down to
a limiting flux of $\rm \sim
10^{-15}~erg~s^{-1}cm^{-2}$), but undetected
at 850$\mu$m: one of these is an AGN at z=0.960 and the
other is an extremely red object whose photometric redshift is estimated
to be z=2.6--2.7 and suspected to host a heavily obscured AGN.
It should be noted that the Chandra observation of the HDFN
is deeper than that presented by M00.
The behaviour of the source counts down to
such faint fluxes is not known yet. If we extrapolate the M00 logN--logS
slope to $\rm 10^{-15}~erg~s^{-1}cm^{-2}$, we estimate that at this limiting
flux about 85\% 
of the 2--10 keV XRB should be resolved if the background
value given in Vecchi et al. (1999) is adopted; using previous background 
values would result in a still higher resolved fraction.

\section{Comparison with AGN and starburst templates}
In Fig.~1 we also show
the values of $\alpha_{SX}$ as a function of
redshift for various classes of objects.

{\it Type 1 unobscured AGNs.}
NCG 4593 has been chosen as a typical Seyfert 1 galaxy
(data from Bicay et al. 
\cite {Bic} and George et al. \cite{Geo}).
Two optically selected PG quasars (PG1613+658 and PG2130+09)
have been taken as typical radio-quiet quasars (data from 
Lawson \& Turner \cite{Law:Tur},  Haas et al. 
\cite{Haa} and Elvis et al. \cite{Elv}).
They are nearby sources ({\it z}=0.12 and 0.06 respectively) with
hard X-ray luminosities$^2$ of
L$_{2-10 {\rm keV}}$=5$\times$10$^{44}$ and  L$_{2-10 {\rm
keV}}$=8$\times$10$^{43}$ erg s$^{-1}$.
Four quasars at $z>4$ (stars) are shown for comparison with the 
PG quasar predictions (data
from McMahon et al. \cite{McM} and
Kaspi et al. \cite{Kas}).

{\it Type 2 obscured AGNs.}
NCG1068 is the archetype object for the class of Seyfert~2 galaxies as 
a whole and at the same time for the Compton thick subclass. In this object the hard X~rays are
due to reflection from both a cold and a warm mirror
 (Matt et al. \cite{Mat}).
SCUBA maps at 850 $\mu$m and at 450$\mu$m
of NGC1068 were obtained by Papadopoulos \& Seaquist
(\cite{Pap:Sea}) and show a nuclear component associated to the AGN
and a circumnuclear component associated to the starforming activity,
which probably accounts for $\sim$2/3 of the sub-mm emission. The
$\alpha _{SX}$ of NGC1068 is shown in Fig. 1; for this object,
we also show the $\alpha _{SX}$
relative to the nuclear component alone (i.e. excluding the sub-mm flux of the
circumnuclear starburst).
We also show the locus of the Circinus\footnote{Sub-mm and far-IR
data are from Siebenmorgen et al. (\cite{Sie}).} galaxy,
a Seyfert~2 object characterized by a 
reflection-dominated spectrum in the 2-10 keV  range and
by a transmitted component above 10 keV
(N$_H = 5\times 10^{24}$ cm$^{-2}$, Matt et al. \cite{Mat1}).
We further show the expected $\alpha _{SX}$  in the case that the absorbing
column density were an order of magnitude
lower than observed, namely $5\times 10^{23}$ cm$^{-2}$.

{\it Starbursts.}
As an example of a powerful starburst galaxy, we plot Arp 220
(data from Rigopoulou et al. \cite{Rig} and
Iwasawa  \cite{Iwa}).

\section{Discussion}

The thick upper horizontal segment in Fig.~1 gives the $\alpha_{SX}$
of the cosmic background, under the assumption that most of the flux
in both spectral windows
comes from redshifts between 1 and 2 (in analogy with the soft XRB).
The lower thick solid line gives the $\alpha_{SX}$ of 
sources which would contribute 100\% of the 2--10~keV XRB and
only 10\% of the sub-mm background\footnote{$\rm \Delta log(F_{850\mu m}
/F_{5keV})=6.53 \Delta \alpha _{SX}$.}.
The location of the $\alpha _{SX}$ of the various
objects in Fig.~1 with respect to these values constrains
the fraction of either background contributed by the dominant sources of
the other background, under the assumption that the objects discussed 
in this paper are indeed representative of the dominant 
contributors to the  XRB and the  sub-mm background, respectively.

{\it Constraints from the bright hard X-ray sources.}
In Fig.~1 dark blue symbols indicate
X-ray sources brighter than $\rm 5\times 10^{-14}~erg~s^{-1}cm^{-2}$
in the 2--10 keV band. At this limiting flux about
30\% of the 2--10 keV background is resolved
(Ueda et al. 1999, M00,
Fiore et al. 2000b). The subsample includes all of the HELLAS
sources\footnote{The 5--10 keV fluxes of the HELLAS sources
were converted into 2--10 keV fluxes by means of the observed X-ray
slope.}, and one of
the sources in F00, whose de-lensed flux is still in the ``bright''
range. It should be noted that these sources are not biased 
in favour of low absorption, due to the selection criterion 
discussed in Sect.~2; in particular the source taken
from F00 is most likely an absorbed AGN, as inferred from the X-ray
and optical data (see F00 for details). Nonetheless,
the $\alpha _{SX}$ upper limits of such ``bright'' sources are 
inconsistent with the values expected for heavily absorbed AGNs.
Instead, they are consistent with the templates of type 1 or, at most,
moderately absorbed AGNs (N$_H < 5\times 10^{23}$ cm$^{-2}$).
The $\alpha _{SX}$ upper limits, when compared to the index of the background
(Fig.~1), show that the sources are underluminous in the sub-mm by
more than two orders of magnitude, if normalized in the X~rays.
On a most conservative approach, even the extrapolation of the 
$\alpha _{SX}$ upper limits to a redshift of $\sim2$ 
with a Compton-thin template falls short of the required 
sub-mm luminosity by a factor of $\sim 50$. Since the hard X-ray sources
brighter than $\rm 5\times 10^{-14}~ erg~s^{-1}cm^{-2}$ in the 2-10 keV
band make only 30\% of the corresponding background, we estimate that they
contribute less than 0.3/50 = 0.6\% of the sub-mm background.
This result is plagued by the limited statistics, on the one hand;
on the other hand, our estimate is very conservative both in the
use of upper limits and in the extrapolation to high redshifts.

{\it Constraints from the faint hard X-ray sources.}
At the limiting flux of $2.5\times\rm 10^{-15}~erg~cm^{-2}$
achieved by M00 at least 75\% of the 2--10 keV
background is resolved. Our sample also includes the fainter sources 
observed by H00 in the HDFN, at a limiting flux of about
$\rm 10^{-15}~erg~s^{-1}cm^{-2}$. As discussed in Sect.~3, at such
low fluxes the resolved fraction of the 2--10 keV background
is probably larger than 85\%. However, we will conservatively
assume 75\% even after inclusion of the HDFN results.

Light blue symbols in Fig.~1 indicate hard X-ray objects whose 2--10 
keV flux is in the range $\rm  5 \times 10^{-14} > F_{2-10keV} >  10^{-15}
~erg~s^{-1}cm^{-2}$. As mentioned above, only one out of these six 
sources, all observed by SCUBA down to a limiting flux of
$\sim 1$ mJy, was detected at 850$\mu$m. Its $\alpha _{SX}$ is consistent
with a heavily absorbed AGN (N$_H = 5\times 10^{24}$ cm$^{-2}$),
however even here the sub-mm flux falls short by a factor of 10
of the luminosity required to match the background colour. 
This only detection, together with the upper limits of the brighter
X-ray sample, tentatively suggest that at lower X-ray fluxes and
at higher redshifts one is observing more heavily absorbed AGNs:
indeed, the locus of the mentioned objects in Fig.~1 cuts across
the templates from the Compton-thin to the Compton-thick ones. The
locations of the SCUBA-undetected, X-ray faint objects are compatible with
the suggested trend, however they are only upper limits: their position
is mainly due to the X-ray data being much deeper for the X-ray weak
sample, at a comparable sub-mm depth.

To estimate the sub-mm contribution of the fainter X-ray sources
we adopt a conservative approach and treat their sub-mm upper
limits as if they were {\it detections}.
We find that this class of sources has an average sub-mm flux (relative
to the 2--10 keV emission) which is less than 9\% of that required by the
sub-mm background. Since, as discussed above,
the X-ray sources in the lower flux range make between 45\% and 70\%
of the 2--10 keV background (the first 30\% being accounted for by the
brighter sources), in the most favourable case they contribute for less than
0.09$\times$0.7$=$6\% to the sub-mm background. Combining this result with the
limit on the brighter sources, we can state that the sources making
the first 75\% of the 2--10 keV background (and perhaps more than
75\%, M00)
contribute no more than 6\%+0.6\% $\sim$ 7\% of the sub-mm background.

We have already cautioned that the limited statistics is a potential problem,
since only 11 hard X-ray sources are used in this study. However, 
all of them but one are upper limits, and the 
{\it censored} Kaplan-Meier estimator gives a very low probability ($<$1\%) 
that the population of sources represented by our sample contribute to the
sub-mm background for a fraction larger than estimated above (more precisely, 
this is the combined probability the mean of the true distribution of 
$\alpha _{SX}$ is higher than 0.89 and 1.04 for the bright and faint sources,
respectively, which correspond to 1\% and 9\% of the background value$^4$).

{\it Constraints from the SCUBA sources.}
The SCUBA sources in our sample have 850$\mu$m (de-lensed) fluxes down
to $\sim$1 mJy. At this limit $\sim$70\% of the sub-mm background is resolved
(Blain et al. 2000). Out of a total of 24, 23 sources do not have
X-ray counterparts in the 2--10 keV band, down to a limiting flux 
of $F_{2-10keV} \sim 1-2\times 10^{-15}$  erg~s$^{-1}$cm$^{-2}$
for most of them. Their lower limits on $\alpha _{SX}$
occupy the upper part of the plot and are
presumably akin to the starburst template given by Arp220. Most of the
SCUBA sources with known redshift
are inconsistent even with the reflection-dominated template given by the
nucleus of NGC1068. These sources are probably dominated by starburst
activity. A significant contribution ($\ge$50\%)
from an obscured AGN might be present {\it if} the latter is completely
Compton thick (i.e. $\rm N_H>10^{25}~cm^{-2}$) {\it and if} the reflection
efficiency is significantly lower than estimated for NGC1068 ($\sim$
1\%). Most of the sources with unconstrained redshift might be
consistent with the NGC1068 template, but are hardly consistent with
AGNs templates which are not reflection-dominated, unless located at very high
redshifts (z$>$3--5). This is in conflict with recent findings according
to which most of the SCUBA faint sources are located at z$<$3 (Barger et al. 1999b).
The presence among the SCUBA-detected sources of AGNs dominated by direct X-ray 
emission seems very unlikely.

\section{Conclusions}

By means of new SCUBA observations and data in the literature we 
constrained the sub-mm emission of hard X-ray (2--10 keV) sources and,
vice versa, the 2--10 keV emission of 850$\mu$m SCUBA sources, at limiting
fluxes which resolve most ($>$70\%) of the cosmic background in the two bands,
i.e. $\sim \rm 10^{-15}~erg~s^{-1}cm^{-2}$ in the 2--10 keV band and
$\sim 1$ mJy at 850$\mu$m.

Only one out of 11 hard X-ray 2--10 keV sources is detected at 850$\mu$m.
This is a Chandra source whose optical, X-ray and sub-mm properties suggest
a type 2, heavily absorbed QSO at redshift between 1.9 and 2.7.
The upper limits on the sub-mm emission of the other X-ray sources are much
lower than the background requirements. In particular, we estimate that, 
under conservative assumptions, the
2--10 keV sources brighter than $\sim \rm10^{-15}~erg~s^{-1}cm^{-2}$, which
resolve at least 75\% of the background in this band, cannot contribute
for more than 7\% to the sub-mm background. This result confirms and
strengthens similar conclusions obtained by F00 and H00.
Any significant contribution to the sub-mm background is limited to
fainter hard X-ray sources which might contribute no more than 25\%
of the 2--10 keV background. 
 These fainter sources should have a sub-mm to X-ray ratio
substantially higher than the stronger ones, of the order
of at least (50/25)/(7/75) $\sim$ 20, if they were to 
contribute 50\% -- say -- of the sub-mm background.
Although the hard X-ray sources which make
most of the 2--10 keV background contribute little to the sub-mm
background, they might contribute significantly to the mid-- and far--IR
background (20--200$\mu$m), since AGN-powered systems are generally
characterized by warmer dust, with a spectral energy distribution peaking
in this range.

None of the 24 SCUBA sources, but the one discussed above, is detected in
the 2--10 keV band down to a limiting flux of $\rm F_{2-10 keV}
\sim 1-2\times 10^{-15}~erg~s^{-1}cm^{-2}$. The lower limits on the ratio
$\rm F_{850\mu m}/F_{5keV}$ indicate that most of them are either powered
by starburst activity or by obscured AGNs which are completely
Compton thick ($\rm N_H > 10^{25}~cm^{-2}$) and, 
in most cases, with an X-ray reflection
efficiency significantly lower than $\sim$1\%.
However, this class of AGNs is not expected 
to contribute significantly to the X-ray background
(e.g. Gilli et al. 1999). Therefore, it remains
true that the sources making most of the sub-mm background do not contribute
significantly to the X-ray background.

Finally, we should mention that most of the 2--10 keV data
used in this work were obtained with Chandra, whose effective area
drops rapidly at high energies, although the sensitivity
is still much higher than previous hard X-ray missions. 
As a consequence, Chandra detections might still be
biased in favor of soft sources dominated by 
photons at about 2 keV. Observations with XMM, which has a much larger
effective area at high energies, should help to tackle this issue.

\begin{acknowledgements}
We are grateful to M. Bolzonella for her help with the photometric redshifts,
and to the JCMT staff for assistance during the observations.
This work was partially supported by the
Italian Space Agency (ASI) through the grant ARS-99-75 and by the
Italian Ministry for University and Research (MURST) through the grant
Cofin-98-02-32.
\end{acknowledgements}

\end{document}